\documentclass{article}

\usepackage{arxiv}

\usepackage[utf8]{inputenc} 
\usepackage[T1]{fontenc}    
\usepackage{authblk}        

\usepackage{amsmath,amsfonts}
\usepackage{url}            
\usepackage{booktabs}       
\usepackage{amsfonts}       
\usepackage{nicefrac}       
\usepackage{microtype}      
\usepackage{lipsum}		
\usepackage{graphicx}
\usepackage{cite}
\usepackage{xcolor} 
\usepackage{etoolbox}
\usepackage[colorlinks,
            linkcolor=red,       
            anchorcolor=green,  
            citecolor=green,        
            ]{hyperref}

\usepackage{natbib}
\usepackage{doi}
\usepackage{gensymb}
\usepackage{CJKutf8}
\usepackage{blindtext}
\usepackage{fancyhdr}
\usepackage{bm}

\makeatletter
\renewcommand\@makefnmark{}
\renewcommand\@makefntext[1]{#1}
\makeatother

\title{DaYu: Data-Driven Model for Geostationary Satellite Observed Cloud Images Forecasting}

\author[1,2]{{Xujun Wei}}
\author[1,2,*]{Feng Zhang}
\author[1,2]{Renhe Zhang}
\author[1]{Wenwen Li}
\author[1]{Cuiping Liu}
\author[1]{Bin Guo}
\author[1]{Jingwei Li}
\author[1,3]{Haoyang Fu}
\author[1,2]{Xu Tang}

\affil[1]{Key Laboratory of Polar Atmosphere-ocean-ice System for Weather and Climate of Ministry of Education / Shanghai Key Laboratory of Ocean-land-atmosphere Boundary Dynamics and Climate Change, Department of Atmospheric and Oceanic Sciences \& Institutes of Atmospheric Sciences, Fudan University, Shanghai 200438, China}
\affil[2]{Integrated Research on Disaster Risk International Centre of Excellence, Fudan University, Shanghai 200438, China}
\affil[3]{College of Physics and Electronic Information Engineering, Zhejiang Normal University, Jinhua 321004, China}

\hypersetup{
pdftitle={DaYu: Data-Driven AI Model for Brightness Temperature Forecasting},
pdfsubject={q-bio.NC, q-bio.QM},
pdfauthor={Xujun Wei, Elias D.~Striatum},
pdfkeywords={First keyword, Second keyword, More},
}

\fancypagestyle{plain}{
    \fancyhead[C]{DaYu: Data-Driven AI Model for Cloud Images Forecasting} 
}

\begin{document}
\maketitle

\footnotetext[1]{* Corresponding author: \texttt{fengzhang@fudan.edu.cn}}

\begin{abstract}
In the past few years, Artificial Intelligence (AI)-based weather forecasting methods have widely demonstrated strong competitiveness among the weather forecasting systems. 
However, 
these methods are insufficient for high-spatial-resolution short-term nowcasting within 6 hours, which is crucial for warning short-duration, mesoscale and small-scale weather events. Geostationary satellite remote sensing provides detailed, high spatio-temporal and all-day observations, which can address the above limitations of existing methods. 
Therefore, this paper proposed an advanced data-driven thermal infrared cloud images forecasting model, "DaYu." Unlike existing data-driven weather forecasting models, DaYu is specifically designed for geostationary satellite observations, with a temporal resolution of 0.5 hours and a spatial resolution of 0.05$\degree$ $\times$ 0.05$\degree$. 
DaYu is based on a large-scale transformer architecture, which enables it to capture fine-grained cloud structures and learn fast-changing spatio-temporal evolution features effectively.
Moreover, its attention mechanism design achieves a balance in computational complexity, making it practical for applications.
DaYu not only achieves accurate forecasts up to 3 hours with a correlation coefficient higher than 0.9, 6 hours higher than 0.8, and 12 hours higher than 0.7, but also detects short-duration, mesoscale, and small-scale weather events with enhanced detail,
effectively addressing the shortcomings of existing methods in providing detailed short-term nowcasting within 6 hours. Furthermore, DaYu has significant potential in short-term climate disaster prevention and mitigation.
\end{abstract}


\vspace{0.3cm}
\section{Introduction}
\label{sec:introduction}

Numerical Weather Prediction (NWP) primarily serves as the foundation for traditional weather forecasting \citep{nwp}, which employs mathematical and physical models to forecast future atmospheric states. 
The main process of NWP involves fusing collected observational data to generate an initial field as close as possible to the actual atmospheric state. This is followed by solving atmospheric dynamics equations and thermodynamic equations to simulate the evolution of the atmosphere. 
Starting from the initial condition, numerical methods are used to step forward in time, generating forecast results.
However, the complexity in solving these equations engenders computational inefficiencies \citep{nwp_1}, significantly limiting the timeliness of weather forecast. 
In addition, 
accurately describing and solving some complex physical processes, such as those in cloud microphysics, using formulas in NWP is highly challenging \citep{nwp_cloud}.
These limitations prompt researchers to explore alternative methods that improve computational efficiency and enhance the forecasting of cloud elements.

Driven by large amounts of data, AI-based weather forecasting systems can provide faster inference speeds and higher accuracy than traditional NWP.
In recent years, researchers have begun exploring AI-based weather forecasting methods. 
\citep{weatherbench} firstly used ResNet to forecast with a spatial resolution of 5.625 $\degree$ for up to 5 days. 
The FourCastNet \citep{fourcastnet}, leveraging the Adaptive Fourier Neural Operator  \citep{afno} and Vision Transformer \citep{vit}, delivers a global weather forecast for 7-day at a spatial resolution of 0.25$\degree$.
These methods can achieve forecast accuracy comparable to traditional NWP.
Pangu-Weather \citep{pangu}, as the first data-driven AI weather forecast model with accuracy surpassing traditional NWP, demonstrates excellent performance in 7-day forecasts at a spatial resolution of 0.25$\degree$. It marks the beginning of a new era in the research of AI-based weather forecasting methods. Additionally, GraphCast \citep{graphcast} implements an autoregressive model based on Graph Neural Networks, outperforming HRES in 90\% of the variable and lead time combinations for 10-day predictions. 
FengWu \citep{fengwu} and FuXi \citep{fuxi} achieve effective forecast duration of at least 10 days, with a maximum reach of 15 days, while further enhancing forecast accuracy. NeuralGCM \citep{neuralgcm} incorporates physical constraints to ensure physically plausible predictions and can be extended to climate simulations.
These methods have demonstrated the potential of data-driven AI-based methods in weather forecasting with their fast inference speed and high accuracy.

Although these AI-based methods represent a significant advancement in weather forecasting, they still suffer some limitations.
The inputs of them are primarily reanalysis data, such as ECMWF Reanalysis v5 (ERA5) \citep{era5}.
Reanalysis data assimilates observational data from a past period with NWP outputs, which leads to delayed timeliness.
More importantly, the temporal resolution of existing AI-based methods is 6 hours, which is insufficient for short-term nowcasting. 
Within the 6-hour time window, there are short-lived convective weather events, which are often reflected by the formation and development of clouds. Existing methods are unable to accurately forecast clouds in the short term, and therefore cannot effectively monitor and warn of short-lived convective weather disasters.
Additionally, the spatial resolution of existing AI-based methods is 0.25$\degree$ $\times$ 0.25$\degree$. As a result, these methods cannot accurately capture the mesoscale and small-scale weather events. And mesoscale and small-scale weather events typically involve specific cloud structures and systems.
To further enhance the forecasting capabilities for short-lived mesoscale and small-scale weather system, it is necessary to design an AI-based short-term forecasting model using high-frequency real-time observational data. 

Data-driven methods using observations from geostationary satellite imager show great promise. 
First, 
geostationary satellite imager observation, with its advantages of high spatio-temporal resolution and wide coverage, enable continuous observing for rapidly changing events such as such as mesoscale convective systems and typhoons.
Based on satellite cloud images, meteorological operational departments can continuously track the position and intensity of tropical cyclones via  Dvorak technique \citep{dvorak,adt}, and monitor mesoscale convective weather systems \citep{convective_1,convective_2}, providing important information for disaster warnings, maritime shipping safety and so on.
In addition, the thermal infrared cloud images of the imagers enable continuous all-day, allowing geostationary satellites to monitor weather changes in a certain region consistently.
Infrared cloud images use thermal infrared radiation to depict cloud top temperatures, thereby characterizing the structure and height of clouds \citep{letu,lww_1,zzj,ljw}.
The geostationary satellite observation data-driven methods can help address the shortcomings of existing AI weather forecast methods which struggle to capture short-term mesoscale and small-scale weather events.
\citep{thunderstorm} uses cloud images at the 10.8 ${\mu}m$ wavelength observed band for short-term thunderstorm forecasting, but this can only achieve four-hour forecasts for small areas. \citep{solar_diffusion} has implemented a combination of brightness temperature and solar radiation forecasting, but the observation area is limited to within 1$\degree$ $\times$ 1$\degree$, resulting in limited coverage. Clearly, existing forecasting models based on satellite remote sensing observations cannot meet the requirements for high spatio-temporal resolution forecasting.

Based on the insights mentioned above, we have innovatively proposed a satellite remote sensing data-driven AI cloud images forecasting model, \textbf{DaYu}. The name "DaYu" (Chinese:
\begin{CJK*}{UTF8}{gbsn}
大禹
\end{CJK*}
) is derived from the legendary story of Dayu controlling the floods in ancient China, which tells of a great case of water management engineering initiated by Yu. Naming this AI model DaYu symbolizes the prevention of meteorological natural disasters.
The DaYu model is a transformer-based autoregressive cascaded model, with each cascaded model predicting time spans of 0-6 hours and 6-12 hours, respectively.
Specifically, this study selected the whole observational data from the Advanced Himawari Imager (AHI), with a temporal resolution of 0.5 hours and a spatial resolution of 0.05$\degree$ $\times$ 0.05$\degree$. 
Following extensive experimental validation, DaYu has demonstrated the capability for accurate 12-hour cloud images forecasting. DaYu can also effectively capture mesoscale and small-scale extreme weather events. This fully demonstrates DaYu's advanced capabilities in short-term forecasting, effectively addressing the shortcomings of existing AI methods in short-term forecasting of mesoscale and small-scale weather events.



Overall, our contribution to this work can be summarized as
follows:
\begin{itemize}
    \vspace{-2mm}
    \item We have innovatively developed an AI-based model "DaYu" for high temporal and spatial resolution satellite cloud images forecasting, achieving a temporal resolution of 0.5 hours and a spatial resolution of 0.05° $\times$ 0.05°. This addresses the gap in temporal and spatial resolution found in current AI-based weather forecasting methods.
    \vspace{-0mm}
    \item DaYu has demonstrated excellent forecasting performance, achieving a skillful forecast lead time (PCC > 0.7) for up to 12 hours in advance. And DaYu has shown robustness in capturing mesoscale and small-scale extreme weather events and has been capable of enhancing the temporal frequency of skillful forecasts to 10 minutes.

\end{itemize}

\section{Preliminary}
\label{sec:preliminary}


\subsection{Data Source}

Himawari Satellite (launched by Japan in 2014) is a new-generation geostationary meteorological satellite. Advanced Himawari Imager (AHI) on the satellite is a passive imager radiometer covers 80$\degree$E to 160$\degree$W and 60$\degree$N to 60$\degree$S. It provides real-time, full-disk observations with different spatial and temporal resolutions across 16 multispectral channels covering 0.47–13.3 $\mu m$, including 6 VNIR channels and 10 infrared channels. 

In this study, we used full-disk, high spatiotemporal resolution Level-1 data with 
a resolution of 0.05$\degree$ $\times$ 0.05$\degree$ provided by the Japan Meteorological Agency (JMA), selecting data at 30-minute intervals to train the model.
Each image data is composed of 2401 $\times$ 2401 uniform grids with a spatial resolution of 0.05$\degree$ $\times$ 0.05$\degree$. 
Meanwhile, we chose to simulate eight infrared brightness temperature channels of the Himawari satellite's AHI. Specifically, we selected the 9th channel through 16th for simulation. Among these, 9th channel and 10th channel are water vapor channels with central wavelengths of 6.9 µm and 7.3 µm, respectively. 11th channel through 16th channel are longwave infrared channels, with central wavelengths ranging from 8.6 $\mu m$ to 13.3 $\mu m$.
The target of each channel is shown in Table~\ref{var_table}.
We use data from 2016, 2017, and 2018 for training, and 2023 for testing. 
It should be noted that using AHI observational data half-hourly over a 3-year period for training requires a comparable amount of data to the six-hourly ERA5 reanalysis data \citep{era5} over a 39-year period.


\subsubsection{Problem Formulation}
The study is based on Himawari satellite AHI observational data and utilizes AI techniques to develop a reliable high-resolution brightness temperature forecasting system. This system predicts the corresponding regional brightness temperature state for the next 12 hours based on current brightness temperature observations. Formally, we represent the brightness temperature state at time $i$ as a high-dimensional tensor \( X_i \in \mathbb{R}^{C \times W \times H} \), where $C$ denotes the number of brightness temperature observation channels considered in this work, and $W$ and $H$ represent the width and height of the brightness temperatur data grid. We map the latitude-longitude equidistant observed brightness temperatures onto a two-dimensional grid with a resolution of 0.05° in both latitude and longitude. Specifically, $C$ = 8, $W$ = 2401, and $H$ = 2401. The objective is to generate a 12-hour advance forecast set \( \{\hat{X}_{i+1}, \hat{X}_{i+2}, \ldots, \hat{X}_{i+24}\} \), with a temporal interval of half an hour.
where \( \hat{X}_{i+\tau} \) represents the prediction of the brightness temperature state at time $i + \tau$. However, based on experience and similar meteorological forecasting efforts, it is challenging to directly learn the mapping from current brightness temperature observations to the brightness temperature states 12 hours into the future, which would result in significant errors. Therefore, DaYu aims to learn a mapping to predict the data for the next time step, and then uses this next time step's data as input to generate multi-step predictions in an autoregressive manner. That is:

\( \hat{X}_{i+1} \) is predicted first, then it is used as part of the input to predict \( \hat{X}_{i+2} \), and so on, continuing this process up to \( \hat{X}_{i+24} \). This approach leverages the most recent predicted state to inform the next prediction, reducing the accumulation of errors that would otherwise occur with a direct forecast.

\begin{table*}[t]
    \centering
    \caption{Himawari/AHI channel-specific central wavelengths and primary purpose \citep{himawari}.}
    \begin{tabular}{ccc}
        \toprule
        Channel & Central Wavelengths & Primary Purpose\\
        \midrule
        9 & 6.9 & {Mid-tropospheric water vapor, wind, and rainfall} \\
        10 & 7.3 & {Lower tropospheric water vapor, wind, and sulfur dioxide} \\
        11 & 8.6 & {Total water content, cloud phase, dust, sulfur dioxide, and rainfall} \\
        12 & 9.6 & {Total ozone, turbulence, and wind} \\
        13 & 10.4 & {Surface and cloud} \\
        14 & 11.2 & {Sea surface temperature, clouds, and rainfall} \\
        15 & 12.4 & {Total water content, volcanic ash, and sea surface temperature} \\
        16 & 13.3 & {Air temperature, cloud height, and cloud cover} \\
        \bottomrule
    \end{tabular}
    \label{var_table}
    \vspace{-0.3cm}
\end{table*}


\section{Methodology}
\label{sec:methodology}

\subsection{Network Framework}
\begin{figure*}[tp]
\centering
\includegraphics[width=0.85\textwidth]{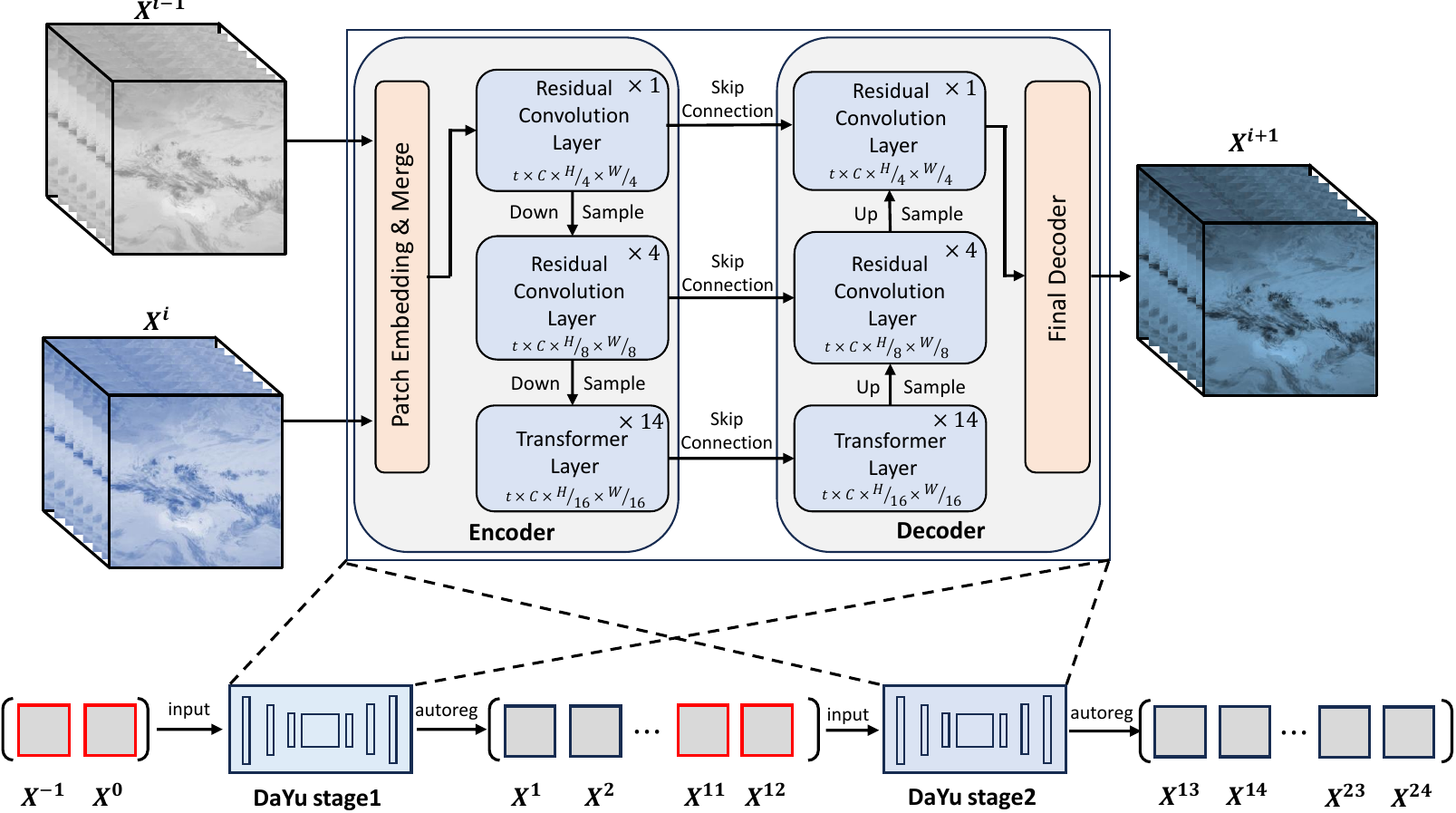}
\caption{Overview of the DaYu Architecture. DaYu cascades the parameters of Stage 1 and Stage 2 models. Taking continuous cloud images from two moments $x^{-1}$, $x^0$ as inputs, DaYu Stage 1 autoregressively forecasts 12 cloud images for the 0-6 hour period, and then uses $x^{11}$, $x^{12}$ as inputs. Stage 2 forecasts 12 cloud images for the 6-12 hour period as the same pattern. Red-bordered boxes indicate the initial inputs for the stage models. Spatio-Temporal Encoder extracts feature embeddings from inputs. Transformer layers are then used to learn the global relationships of high-level abstract features. Finally, Spatio-Temporal Decoder generates the predicted cloud image for the next moment. In this figure, $i$ ranges from 0 to 23.}
\label{framework}
\vspace{-0.4cm}
\end{figure*}

Since the cloud image (brightness temperature) observational data is a multichannel data matrix with spatial continuity between the grid values, 
DaYu
adopted Encoder-Decoder structure, as shown in Figure~\ref{framework}, drawing inspiration from fine-grained image perception tasks in the field of computer vision. 
First, the encoder extracts the features of the cloud image inputs, with the inputs being a 2 $\times$ 8 $\times$ 2401 $\times$ 2401 data, where 2 represents two consecutive time steps, 8 represents the total number of imager observational channels, 2401 $\times$ 2401 means the latitude ($H$) $\times$ longitude ($W$) grid points, respectively. The global features are modeled through a transformer-based module, and finally, the decoder outputs a 1 $\times$ 8 $\times$ 2401 $\times$ 2401 data, which represents the prediction for the next time step.

\textbf{Spatio-Temporal Encoder-Decoder.} \ First, considering that a 8 $\times$ 2401 $\times$ 2401 feature map is too large for current AI models and computational resources, it is necessary to design sufficient feature encoders to enable the model to adequately capture key features at a manageable size. Additionally, reducing the dimension of the feature can accelerate the training of the model. Therefore, the feature encoder employs multiple layers of convolutional neural networks.
To extract features quickly and effectively, we first use patch embedding and merge block to process features from two consecutive time steps. The convolutional patch embedding block not only reduces the temporal and spatial dimensions but also encodes the features from two consecutive time steps into a single fused feature, making the model easier to learn. Specifically, the convolution embedding blocks are composed of convolutions with kernel sizes of 4 and strides of 4. Additionally, since the width and height of the features are 600 after embedding blocks, the computational and memory requirements remain high. Therefore, Spatio-Temporal Feature Encoder also employs two downsampling residual convolution layers with 3 $\times$ 3 convolution kernels to further reduce the dimensionality.

\textbf{Spatial-Affine Transformer.} \ 
Vision Transformers based on multi-head attention have demonstrated strong performance in visual tasks, and some variants \citep{swin_transformer} are widely used in the field of meteorology. However, the Transformer architecture not only requires large amounts of data to train effectively but also needs appropriately sized feature maps as input. After the Residual convolution layers of the Spatio-Temporal Encoder part, the feature map size is 150 $\times$ 150. At these feature resolutions, the computational cost of multi-head attention is very high. Therefore, in the third feature layer, DaYu stacks 24 Transformer modules to fully extract global features, where the form of multi-head attention is : 

\begin{equation}
    \text{Attention}(\text{Q, K, V}) = \left( \text{softmax} \left( {{\text{Q}\phi(\text{K})}^\top}/{\sqrt{D}} + \text{B} \right) \right) \phi(\text{V})
\label{eq:attention}
\end{equation}
where Q, K, and V are the query, key, and value vectors generated by the Transformer block, respectively. D serves as the feature dimension of the Q, K vectors, and B is the position bias term. $\phi(\cdot)$ represents the feature dimension reduction layer. To further adapt to large-scale feature inputs, $\phi(\cdot)$ uses a 6 $\times$ 6 convolutional kernel to sample features. The $\phi$(K) and $\phi$(V) vectors represent the processed reduction feature vectors.
This attention mechanism can significantly reduce computational complexity while ensuring the modeling capability of features.
By using residual connections, the output vectors of the two Transformer layers are concatenated, and then fused into a unified feature representation through an Multi-Layer Perceptron and GeLU activation function, followed by upsampling.



\subsection{Model training}
This section outlines the training process of the research model. The training process consists of two steps: (1) pretraining and (2) multi-step finetuning.

\subsubsection{Pretraining}
The pre-training process uses the training dataset to predict a single time step, during the pre-training process, the current time step's observational cloud image and the observation from the previous time step are used as input, while the cloud image observation from the next time step serve as the labels to supervise the single-step model's prediction for the next time step. The data flow of the pretraining is defined as follows:

\begin{equation}
    \hat{X}^{t+1} = DaYu(X^{t-1}, X^{t}), 
\end{equation}
where $X^{t-1}$ is the data of previous time step $t-1$, $X^{t}$ is the data of current time step $t$, and $\hat{X}^{t+1}$ is the prediction of the next time step $t+1$.
During model training, residual learning is employed to learn the difference between the input and output, which improves prediction accuracy. Specifically, the model focuses on capturing the difference between the actual observed values and the model's predicted values, rather than directly predicting the target variable. This approach helps the model more effectively learn the dynamic relationships between each time step.
Meanwhile, the loss function used is the Mean Squared Error ($MSE$) loss $\mathcal{L}_{MSE}$, which is defined as follows:

\begin{equation}
    \mathcal{L}_{MSE} = \frac{1}{C \times H \times W} \sum_{c=1}^{C} \sum_{h=1}^{H} \sum_{w=1}^{W} (X_{c,i,j}^{t+1} - \hat{X}_{c,i,j}^{t+1})^2, 
\end{equation}
where $X_{c,i,j}^{t+1}$ is the true brightness temperature value for grid point at channel $c$, height $i$, and width $j$, while $\hat{X}_{c,i,j}^{t+1}$ is the corresponding predicted value, $C$ is the number of channels, $H$ is the height size of the data, and $W$ is the width size of the data.

DaYu was developed using the Pytorch framework. The pre-training of the model requires approximately 48 hours on a cluster of 8 Nvidia A100 GPUs. The model is trained for 60,000 iterations with a batch size of 1 on each GPU. The AdamW optimizer is used with parameters $\beta_{1}$ = 0.9 and $\beta_{2}$ = 0.95, an initial learning rate of $1.5 \times 10^{-4}$, and a weight decay coefficient of 0.05. Additionally, Distributed Data Parallel (DDP), bfloat16 floating point precision, and gradient checkpointing are employed to reduce the memory cost during the model training process. 


\subsubsection{Multi-step finetuning}
After pre-training, we need to fine-tune the base pre-trained model to achieve optimal performance for predicting cloud images up to 12 hours. This fine-tuning process uses an auto-regressive training paradigm, gradually increasing the autoregressive steps from 2 to 24, every step is half-hour, following the fine-tuning method similar to previous large weather models \citep{graphcast, fuxi} based on ERA5 reanalysis product. Additionally, to reduce the accumulation error, we adopt staged fine-tuning approach, where the model is fine-tuned and saved within fixed prediction leading time. These sub-models are then combined in a cascading manner to achieve good predict capability.

The final predictions from DaYu take the form of autoregressive predictions, using the initial true observed cloud image as the starting for the autoregressive prediction. Each single-time-step prediction output is used as the input for the next time step in the autoregressive process. This cycle continues until the output for the last time step is produced, which is the predicted cloud image 12 hours later.

\begin{equation}
    \hat{X}^{t+1} = DaYu(X^{t-1}, X^{t}),  \hat{X}^{t+2} = DaYu(X^{t}, \hat{X}^{t+1}), ... , \hat{X}^{t+24} = DaYu(\hat{X}^{t+22}, \hat{X}^{t+23}),
\end{equation}

\subsection{Evaluation method}
Due to the fact that most previous large-scale AI models were based on ERA5 reanalysis data, they used latitude-weighted evaluation metrics. For the observation area of Himawari satellite AHI, we treat each point of the grid equally, adopting the average \textbf{PCC} (Pearson Correlation Coefficient) and \textbf{RMSE} (Root Mean Square Error) as quantitative analysis metrics, which are calculated as follows:

\begin{equation}
\label{rmse_eq}
    \text{RMSE}(c, \tau) = \frac{1}{T}{\sum_{i=1}^{T}}\sqrt{\frac{1}{W \cdot H} {\sum_{w=1}^{W}}{\sum_{h=1}^{H}} (x^{i+\tau}_{c,w,h} - \hat{x}^{i+\tau}_{c,w,h})}
\end{equation}
\begin{equation}
\label{pcc_eq}
    \text{PCC}(c, \tau) = \frac{1}{T}{\sum_{i=1}^{T}}\frac{\sum_{w,h} (x^{i+\tau}_{c,w,h} - \bar{x}^{i+\tau}_{c})(\hat{x}^{i+\tau}_{c,w,h} - \bar{\hat{x}}^{i+\tau}_{c})}{\sqrt{\sum_{w,h} (x^{i+\tau}_{c,w,h} - \bar{x}^{i+\tau}_{c})^2 \sum_{w,h} (\hat{x}^{i+\tau}_{c,w,h} - \bar{\hat{x}}^{i+\tau}_{c})^2}}
\end{equation}
where $T$ represents all initial times, $\tau$ represents the current forecast duration, $x^{i+\tau}_{c,w,h}$ represents the true observed value of central wavelength $c$ at longitude $w$ and latitude $h$, and $\hat{x}^{i+\tau}_{c,w,h}$ represents the forecast value of central wavelength $c$ at longitude $w$ and latitude $h$. 
$\bar{x}^{i+\tau}_{c}$ represents the mean value of the true observed value of central wavelength $c$, and $\bar{\hat{x}}^{i+\tau}_{c}$ represents the mean value of the forecast value of central wavelength $c$.

Note that the PCC (Pearson Correlation Coefficient) used here is different from the Anomaly Correlation Coefficient (ACC) proposed by the European Centre for Medium-Range Weather Forecasts (ECMWF). The ACC is based on the global climatology calculated from ERA5 reanalysis data, which spans a long period of time, and the ACC measure is sensitive to outliers and can be influenced by various factors such as the choice of time period and spatial resolution of the data. Different choices of these factors can lead to significant differences in the ACC \citep{fengwu}. Therefore, in this study, the correlation coefficient calculation for large-scale, high-frequency satellite remote sensing data uses the PCC.





\section{Results}
\label{sec:results}
We report the results of the cloud images (brightness temperature) forecasting. This study uses the 2023 observational data from the Himawari satellite AHI to conduct a comprehensive evaluation of DaYu, and selects two daily initialization times (00:00 UTC and 12:00 UTC) to generate the forecasts for 12 hours.

\subsection{Evaluation Analysis}
\begin{figure*}[t]
\centering
\includegraphics[width=1\textwidth]{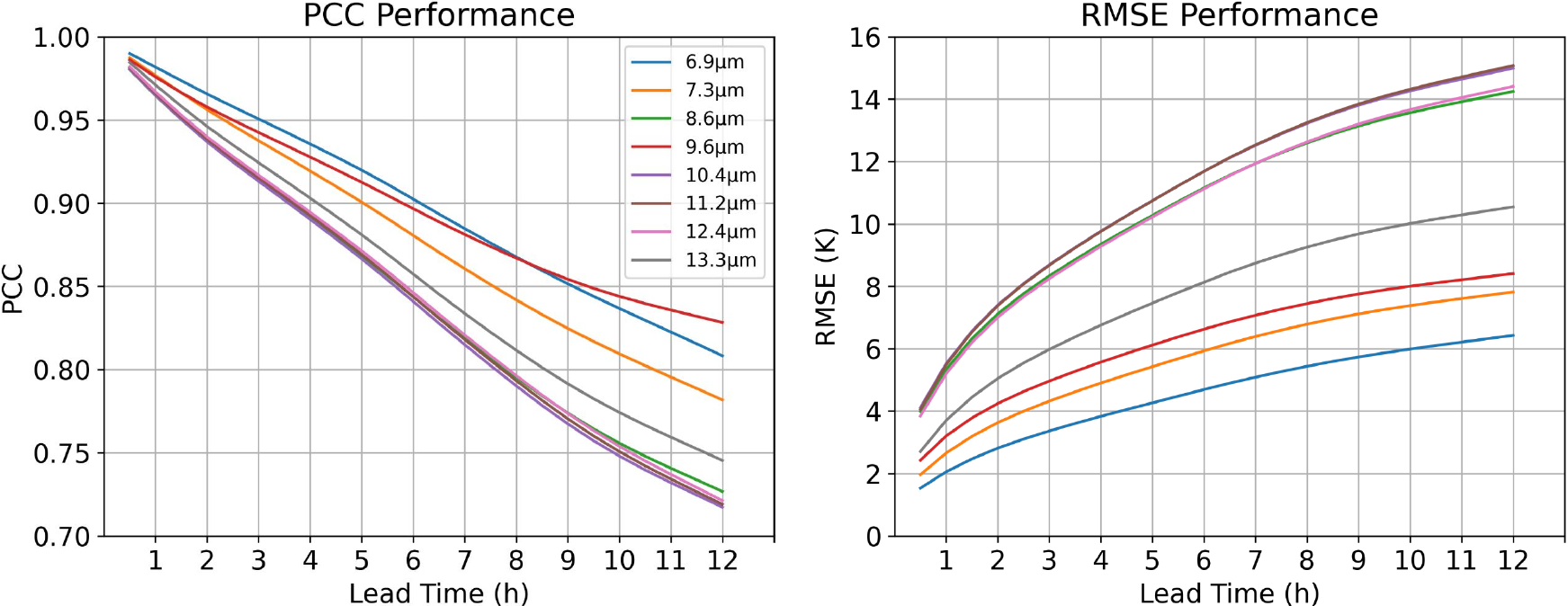}
\caption{\textbf{PCC} and \textbf{RMSE} of DaYu forecasting the brightness temperature states in 2023. In each subplot, the x-axis represents the lead time, with intervals of 0.5 hours over a 12-hour lead time. The y-axis represents the PCC and RMSE as defined in the Eq.~\ref{pcc_eq} and Eq.~\ref{rmse_eq}.}
\label{acc}
\vspace{-0.3cm}
\end{figure*}

Figure~\ref{acc} illustrate the forecast performance of this study in terms of PCC and RMSE, respectively. Each day, two initial times (00:00 UTC and 12:00 UTC) were used, with a forecast leading time of 12 hours for each initial time. The results indicate that for forecasts within3 hours, the PCC can reach above 0.9. For forecasts between 3-6 hours, the PCC can reach above 0.8. For forecasts between 6-12 hours, the PCC can reach above 0.7. 
The channels at 8.6 $\mu m$, 10.4 $\mu m$, 11.2 $\mu m$, and 12.4 $\mu m$ exhibit reduced forecast skill than other channels in their observations. This decreased accuracy can be attributed to their high sensitivity to clouds. These channels detect rapid changes in mesoscale cloud systems, leading to significant variations in the observed cloud images. 
As a result, the frequent and rapid changes in these channels make accurate forecasting more challenging.

\begin{figure*}[t]
\centering
\includegraphics[width=1\textwidth]{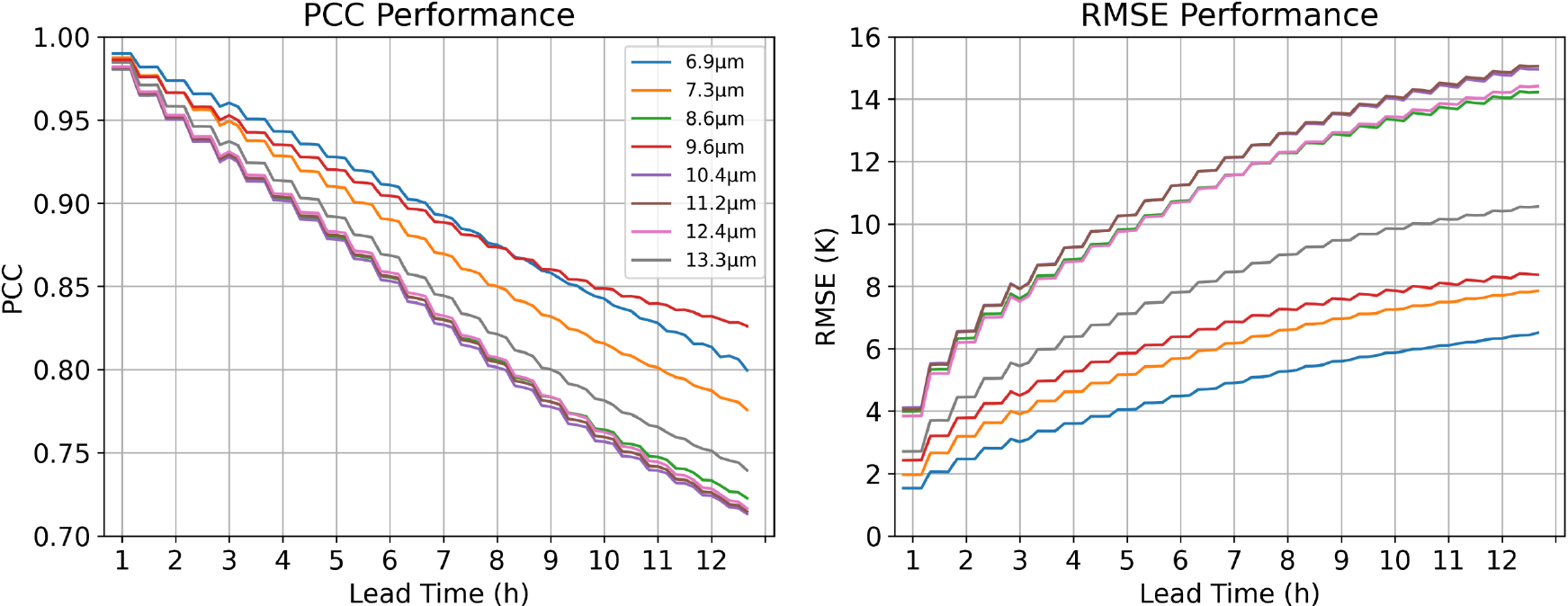}
\caption{\textbf{PCC} and \textbf{RMSE} of DaYu forecasting the cloud image in 2023. In each subplot, the x-axis represents the lead time. The y-axis represents the PCC and RMSE. By arranging the forecast results of three consecutive 10-minute intervals in chronological order, we obtain a forecast result with a temporal resolution of 10 minutes.}
\label{acc_rmse_super}
\end{figure*}

Although DaYu has a temporal resolution of half an hour, it can achieve a time super-resolution effect by using multiple continuous initial time points. Specifically, the conventional input is \( x_{t-30\text{min}}, x_t \), producing outputs for the future interval of 30 minutes across 24 consecutive time points of cloud image, where the $t$ means the time at $t$ o'clock or $t$-thirty. When the input is changed to \( x_{t-20\text{min}}, x_{t+10\text{min}} \) or \( x_{t-10\text{min}}, x_{t+20\text{min}} \), the model can still output results for the future 24 time points (12 hours).
From Figure~\ref{acc_rmse_super}, it can be observed that under the same number of autoregressive forecast steps, the forecast results from three different initial time points are largely consistent, without any uncontrollable fluctuations. Although data's time with minute values of 10, 20, 40, and 50 were not used during training, DaYu's performance in forecasting based on these unseen initial time data still conforms to the patterns shown in the Figure~\ref{acc}. This demonstrates the robustness of DaYu.

\begin{figure*}[t]
\centering
\includegraphics[width=0.95\textwidth]{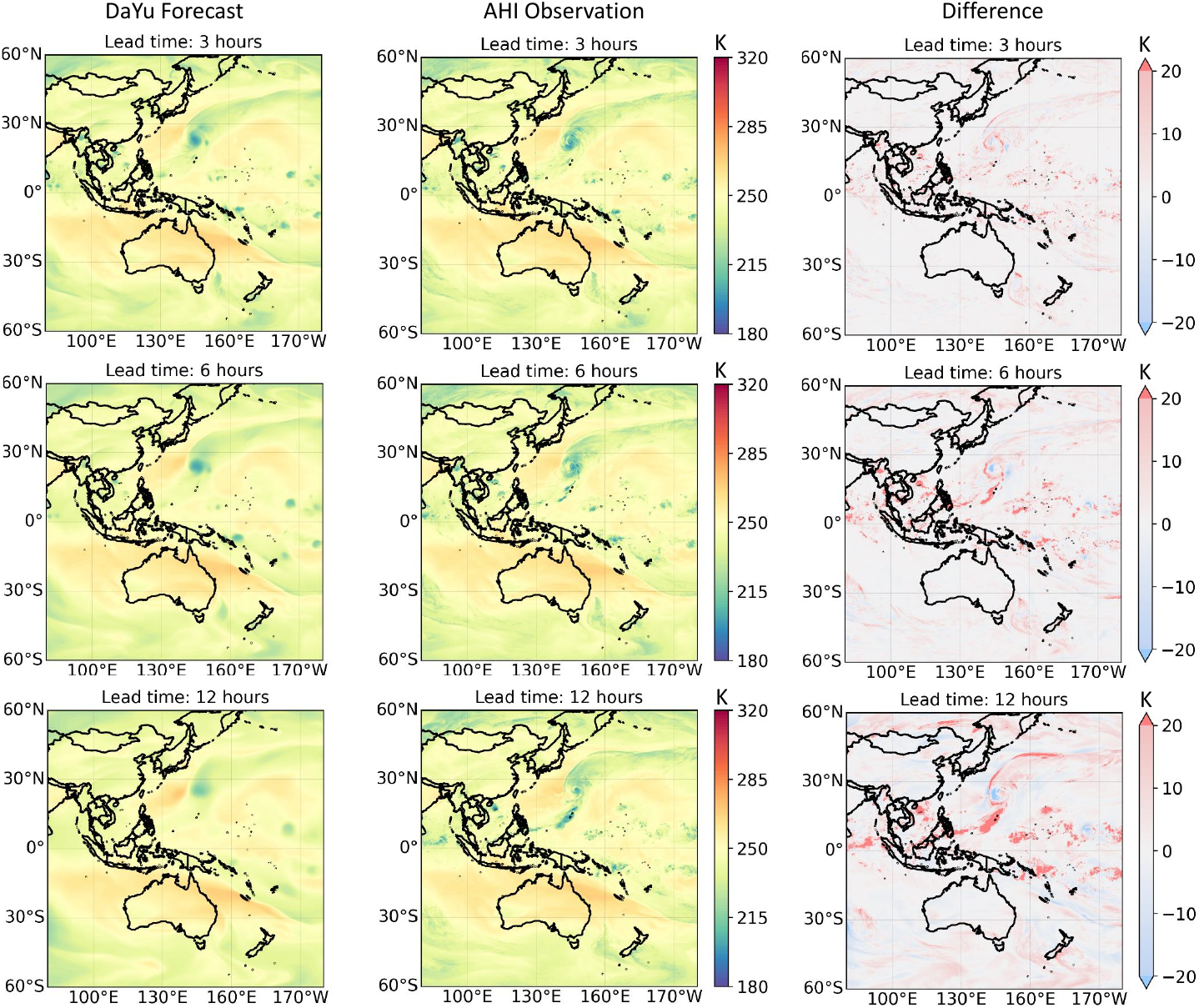}
\caption{\textbf{Visualization of forecast results for the observational wavelength at 6.9\bm{$\mu m$}}. The visualization of cloud image forecasts for 3 hours (first row), 6 hours (second row) and 12 hours (third row). For each row, the results shown are: DaYu forecast (left), AHI observation (middle), and the difference between DaYu forecast and AHI observation (right). For all cases, the initial time is 12:00 UTC on October 12, 2023.}
\label{qualitative_results}
\vspace{-0.5cm}
\end{figure*}

\begin{figure*}[t]
\centering
\includegraphics[width=0.95\textwidth]{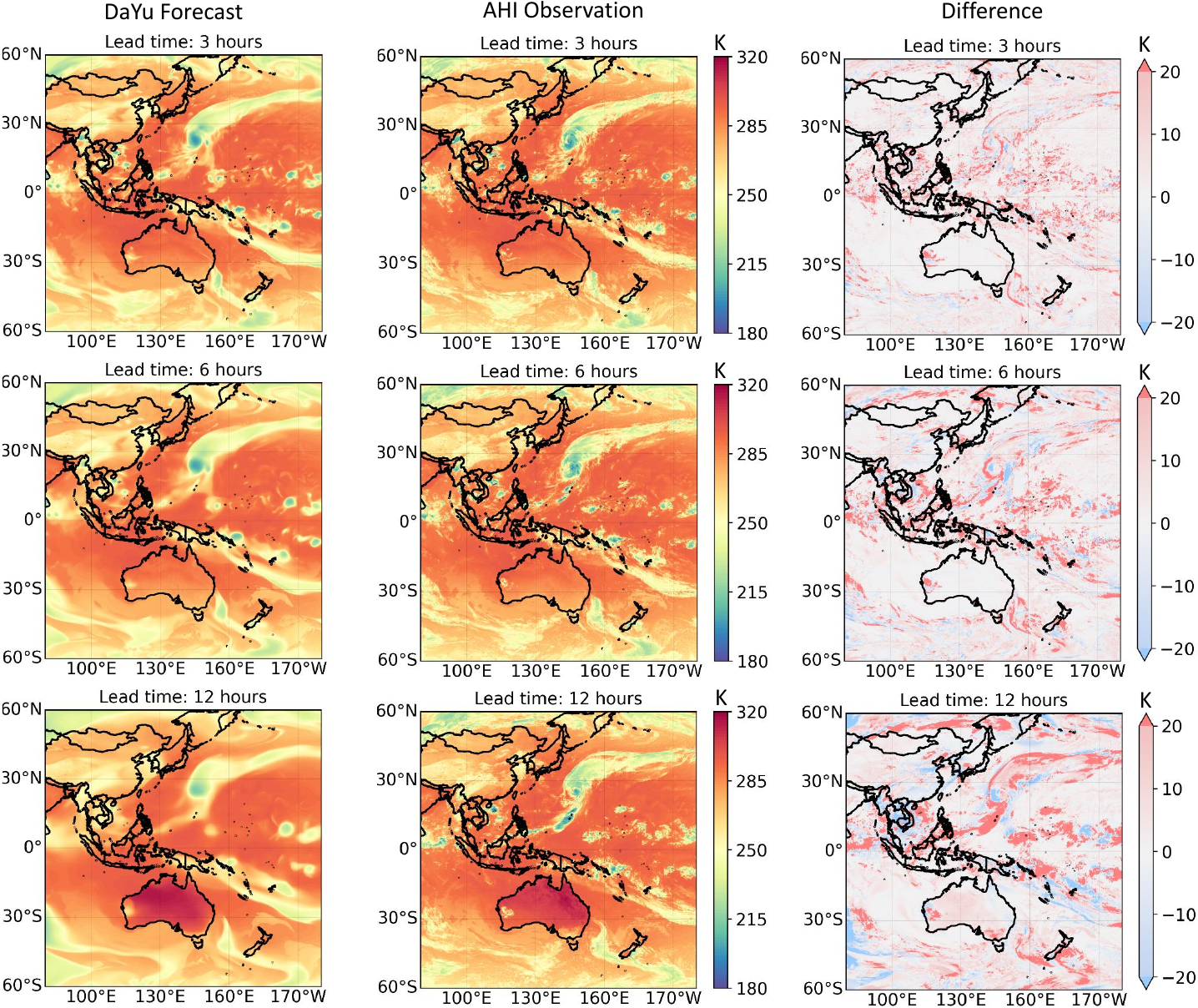}
\caption{\textbf{Visualization of forecast results for the observational wavelength at 11.2\bm{$\mu m$}}. The visualization of cloud image forecasts for 3 hours (first row), 6 hours (second row) and 12 hours (third row). For each row, the results shown are: DaYu forecast (left), AHI observation (middle), and the difference between DaYu forecast and AHI observation (right). For all cases, the initial time is 12:00 UTC on October 12, 2023.}
\label{qualitative_channel_5_results}
\end{figure*}

\subsection{Case Study}
This study selected a case with an initial time of 12:00 UTC on October 12, 2023, 
this initial time falls within the lifecycle of Typhoon Bolaven, and this case can significantly reflect the model's forecast ability to forecast typhoons.
Through comprehensive evaluation using PCC and RMSE, this experiment determines that the 6.9 $\mu m$ central wavelength channel represents the easiest channel for predicting, while the 11.2 $\mu m$ central wavelength channel represents the most challenging channel for the model.
DaYu's forecast of 6.9 $\mu m$ at 3 hour, 6 hours and 12 hours, actual satellite observations and their difference are illustrated in Figure~\ref{qualitative_results}. The model's forecasts closely match the true brightness temperature in most areas. As the forecast steps increase, the error also increases. Figure~\ref{qualitative_channel_5_results} show the comparison on 11.2$\mu m$ between DaYu's forecast and observation. Notably, DaYu performs well in the rapid changing systems such as tropical cyclone and mesoscale convective systems. Differently, the variation of brightness temperature and prediction error on 11.2 $\mu m$ is greatly larger than that on 6.9 $\mu m$.

The 6.9$\mu m$ observational channel, with sensitivity to mid-to-upper atmospheric water vapor and cloud layers, offers limited penetration through lower-level clouds. Primarily capturing the temperature and water vapor of mid-to-high-level clouds, it remains relatively unaffected by low-level clouds and surface temperature variations. This characteristic enables the channel to yield stable and consistent brightness temperature data in the presence of mid-to-upper clouds, allowing models to effectively capture atmospheric middle-layer and cloud structures, thereby reducing forecast errors.
In contrast, the 11.2$\mu m$ observational channel, characterized by strong penetration capabilities, partially observes cloud-top and surface temperatures by passing through thin clouds. Its sensitivity to cloud layers across various altitudes, especially low-level clouds and surface temperatures, introduces greater uncertainty due to the variability in low-level cloud height and thickness and the influence of surface types and diurnal temperature changes. Differences in behavior between cloudy and cloud-free areas further complicate consistent model predictions, often resulting in less stable forecast performance compared to the 6.9$\mu m$ channel.

\begin{figure*}[t]
\centering
\includegraphics[width=1\textwidth]{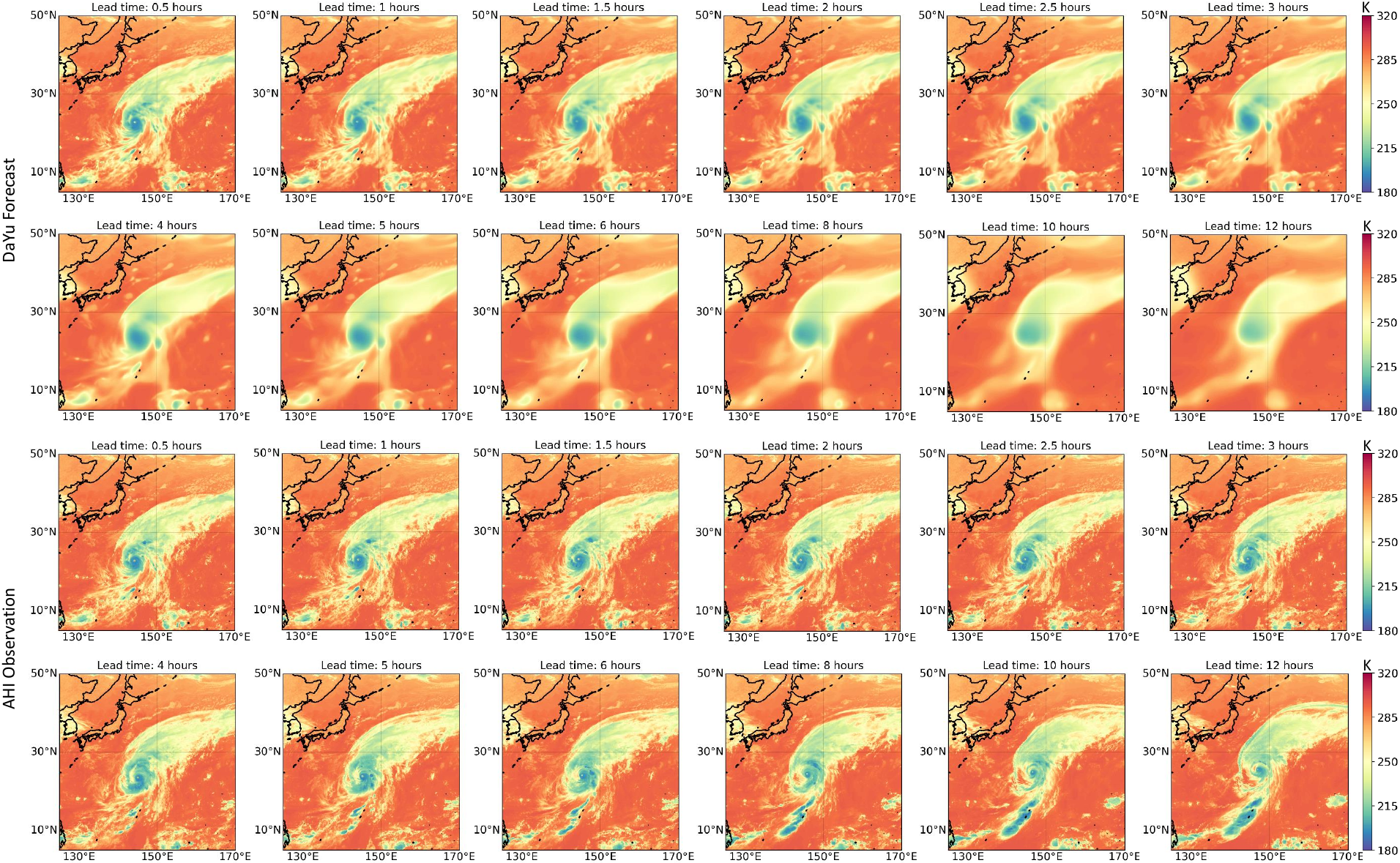}
\caption{\textbf{Visualization of forecast results for Super Typhoon "Bolaven"}. 
The first two rows show the cloud images from DaYu forecasts at different lead times, while the last two rows display the AHI observed cloud images.
For all cases, the initial time is 12:00 UTC on October 12, 2023.
}
\label{typhoon}
\end{figure*}

Figure~\ref{typhoon} shows DaYu's forecast results for the typhoon, using the 11.2 $\mu m$ channel. 
Compared to existing data-driven AI methods, DaYu has a temporal resolution of 0.5 hours, allowing for a higher frequency for capturing the rapid weather systems.
The forecast results within the illustrated region in Figure~\ref{typhoon} were evaluated, yielding the following metrics: the PCC at 3 hours is 0.913 with an RMSE of 5.839; the PCC at 6 hours is 0.869 with an RMSE of 7.079; and the PCC at 12 hours is 0.832 with an RMSE of 9.058.
Within the 0-2 hour period, the model accurately depicts the structure of the typhoon cloud system and the position of the eye, which has great application potential on typhoon nowcasting.

\section{Conclusions and Discussions}
\label{sec:discussions}
Currently, AI-based weather forecasting methods have achieved significant breakthroughs, but they typically use reanalysis data with a 6-hour interval and a spatial resolution of 0.25$\degree$ $\times$ 0.25$\degree$ for training and inference. This limitation means that these methods cannot achieve end-to-end forecasting. More importantly, a 6-hour time interval is insufficient for capturing short-lived severe convective systems events, and a spatial resolution of 0.25$\degree$ $\times$ 0.25$\degree$ kilometers may overlook mesoscale and smaller-scale weather. 
Satellite remote sensing, as a real-time observation technology, can provide large-scale, high-frequency, and detailed observations, addressing the limitations of reanalysis data from a data perspective. Therefore, this paper introduces a satellite infrared brightness temperature forecasting method. We innovatively developed an AI-based model \textbf{DaYu} for cloud images forecasting with a temporal resolution of 0.5 hours and a spatial resolution of 0.05$\degree$ $\times$ 0.05$\degree$. DaYu can provide skillful cloud image forecasts up to 12 hours, overcoming the shortcomings of existing models that lack short-term nowcasting capabilities within 6-hour intervals. It has significant potential in regional climate disaster prevention and mitigation.

Despite the fact that DaYu has achieved skillful 12-hour infrared cloud image forecasts, a comprehensive evaluation of its application potential requires further downstream application studies. Specifically, the potential value can be assessed by by leveraging the correlation between cloud products, water vapor data, and the infrared cloud images. These correlations not only help in validating the accuracy of the model's forecasts but also open up new application for the model. Additionally, future work can explore how to generate more detailed and refined forecasts to meet the needs of different scenarios.


\section{Acknowledgements}
\label{acknowledgements}
We acknowledge the Level 1 cloud image observational data from the Himawari AHI provided by the Japan Meteorological Agency (JMA), which was crucial for this research. We acknowledge the support of National Natural Science Foundation of China (No.42222506).
Additionally, the computations in this research were performed using the CFFF platform of Fudan University. With the support of these resources, the full content of this study was made possible.

\bibliographystyle{unsrtnat}
\bibliography{references}  






\end{document}